\documentstyle[iopconf1,psfig]{article}
\begin{document}
\title{Searching~New~Physics~in~Muonium~Atoms}
\author{Klaus P. Jungmann\footnote{E-mail:
jungmann@physi.uni-heidelberg.de}}
\affil{Physikalisches Institut, Universit\"at Heidelberg, Philosophenweg~12,
D-69120 Heidelberg, Germany}
\beginabstract
Quantum Electrodynamics describes electromagnetic interactions 
of free muons and of muon and electron in the muonium atom to very high 
accuracy.
Contributions of other and yet unknown interactions can be investigated in 
precision experiments, e.g. by searching  
for  muonium-antimuonium conversion and by measuring 
the muon magnetic anomaly. 
\endabstract

\section{Muonium-antimuonium conversion}
The hydrogenlike muonium atom ($M= \mu^+ e^-$) consists of two leptons from
different generations.
The close confinement of the bound state offers excellent opportunities to 
explore precisely fundamental electron-muon interaction. The dominant 
part of the binding in this atom is electromagnetic 
which can be calculated to very high 
accuracy in the framework of quantum electromagnetics (QED). Indeed, precision
experiments of electromagnetic transitions in this system have been employed
both for testing the validity of bound state QED calculations 
and for determining
most accurate values of fundamental constants \cite{Hughes_90}. 
The present 36~ppb precision of measurements 
of the hyperfine structure of muonium is already a factor of two below the 
50~ppb
contribution of strong interactions arising from hadronic vacuum polarization
New experimental results on this transition from recently 
completed measurements at
the Los Alamos Meson Physics Facility (LAMPF) are expected to reveal a 15~ppb
contribution of weak interaction due to $Z_0$ boson exchange. 

\begin{figure}[t]
\label{theo_mmb}
\unitlength 1.0 cm 
 \begin{picture}(11.0,3.8)
  \centering{
   \hspace{0.9 cm}
   \raisebox{-0.2 cm}{
   \mbox{
   \psfig{figure=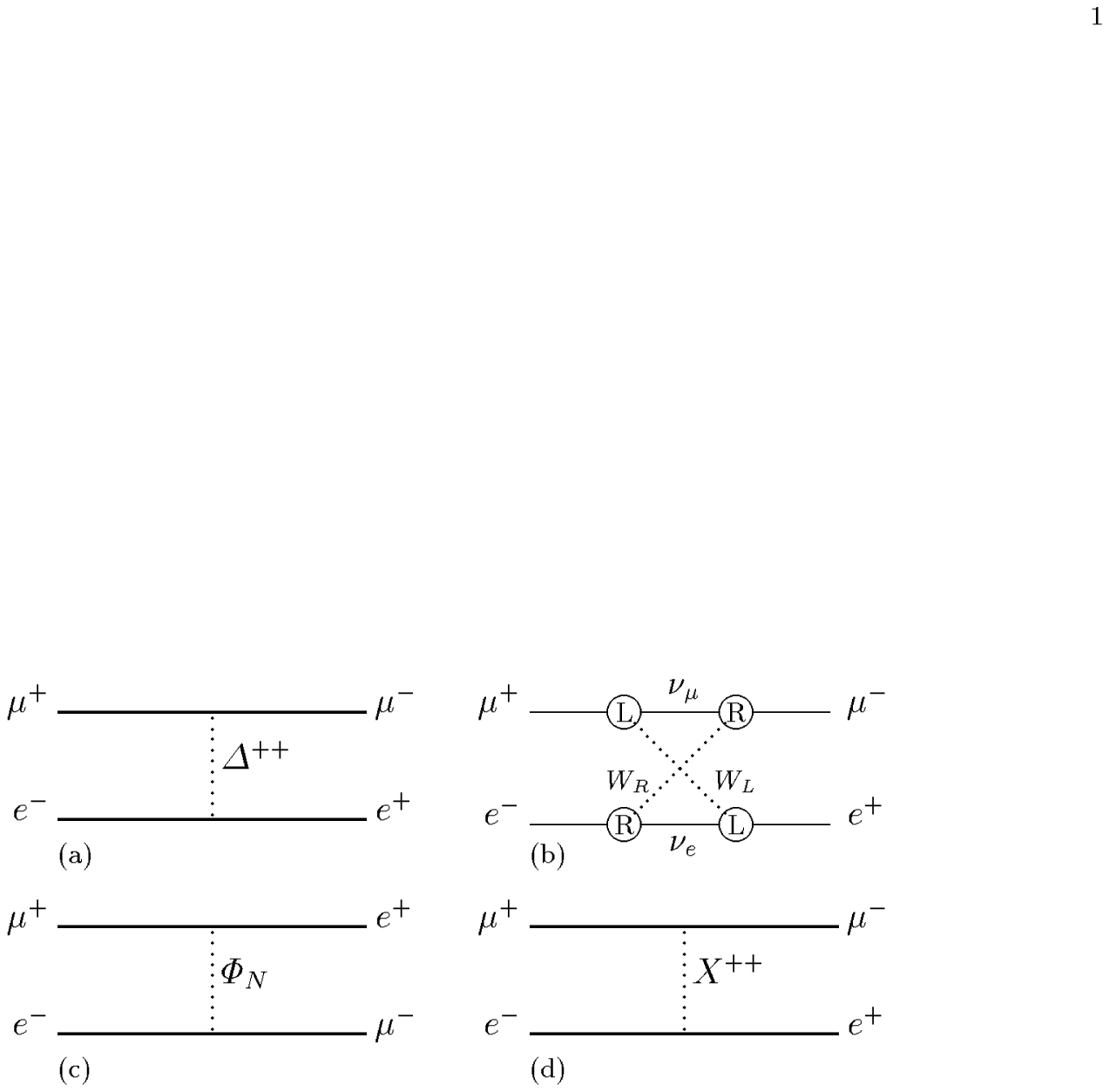,height=10cm}
        }
        }
             }
 \end{picture} 
 \centering\caption[]
        {
        Muonium-antimuonium conversion in 
        theories beyond the standard model. The interaction
        could be mediated by
        (a) a doubly charged Higgs boson $\Delta^{++}$ 
        \protect{\cite{Herczeg_92}},
        (b) heavy Majorana neutrinos \protect{\cite{Halprin_82}},
        (c) a neutral scalar $\Phi_N$ \protect{\cite{Hou_96}}, e.g.
        a supersymmetric $\tau$-sneutrino $\tilde{\nu}_{\tau}$
        \protect{\cite{Mohapatra_92}}, or
        (d) a dileptonic gauge boson $X^{++}$ \protect{\cite{Sasaki_94}}.
        } 
\end{figure}

Since the effects of all known fundamental forces in this system are calculable 
very well, it renders the possibility to search sensitively for yet unknown 
interactions between both particles. A conversion of muonium into
its antiatom ($\overline{M} = \mu^- e^+$) would violate additive lepton family 
number conservation. Although no symmetry behind this experimentally 
established law has been found, 
the process is not provided in standard theory.
However, muonium-antimuonium conversion appears 
to be natural in many speculative 
theories (see Fig. \ref{theo_mmb}), 
which try to extend the standard model in order to explain 
some of its yet not well understood features like parity violation in 
weak interaction and particle mass spectra. 
An experiment has been set up to search for spontaneous 
muonium-antimuonium conversion at the Paul Scherrer Institute (PSI) 
in Villigen, Switzerland (Fig. \ref{mmbarsetup}) \cite{Abela_96}. 
It uses the powerful signature developed 
in an experiment at LAMPF, which requires the coincident identification
of both particles forming the antiatom in its decay 
\cite{Matthias_91,Willmann_97}.

Muonium atoms were produced by stopping a beam of "surface" muons 
in a SiO$_2$ powder target, where a 60 \% fraction of them forms
muonium atoms by electron capture some 5\% of which diffuse through 
the target surface
with thermal energies. Energetic electrons from the decay of 
the $\mu^-$ in the antiatom can be observed in a magnetic spectrometer
at 0.1~T magnetic field
consisting of five concentric multiwire proportional chambers 
and a 64 fold segmented hodoscope. 
The positron in the atomic shell of the antiatom
is left behind after the decay with 13.5~eV average kinetic energy.
It can be accelerated to 8~keV in a two stage electrostatic device and guided 
in a magnetic transport system onto a position sensitive microchannel
plate detector (MCP). Annihilation radiation can be observed in a 12 fold 
segmented pure CsI calorimeter surrounding the MCP.
\setcounter{figure}{2}
\begin{figure}[bt]
\protect{\label{mmbarsetup}}
\unitlength 1.0cm
\begin{minipage}{2.95cm} 
{{\bf Figure 2.}
\sloppy
              Top view of the
              apparatus to search for
              mu\-oni\-um--an\-ti\-mu\-oni\-um con\-vers\-ion
              at PSI \protect{\cite{Abela_96}}.
} 
\end{minipage}
\hfill 
\unitlength 1.0cm 
\begin{minipage}{7.6cm}
\begin{picture}(5.5,6.6)
\hspace{0.0 cm}
\mbox{
\psfig{figure=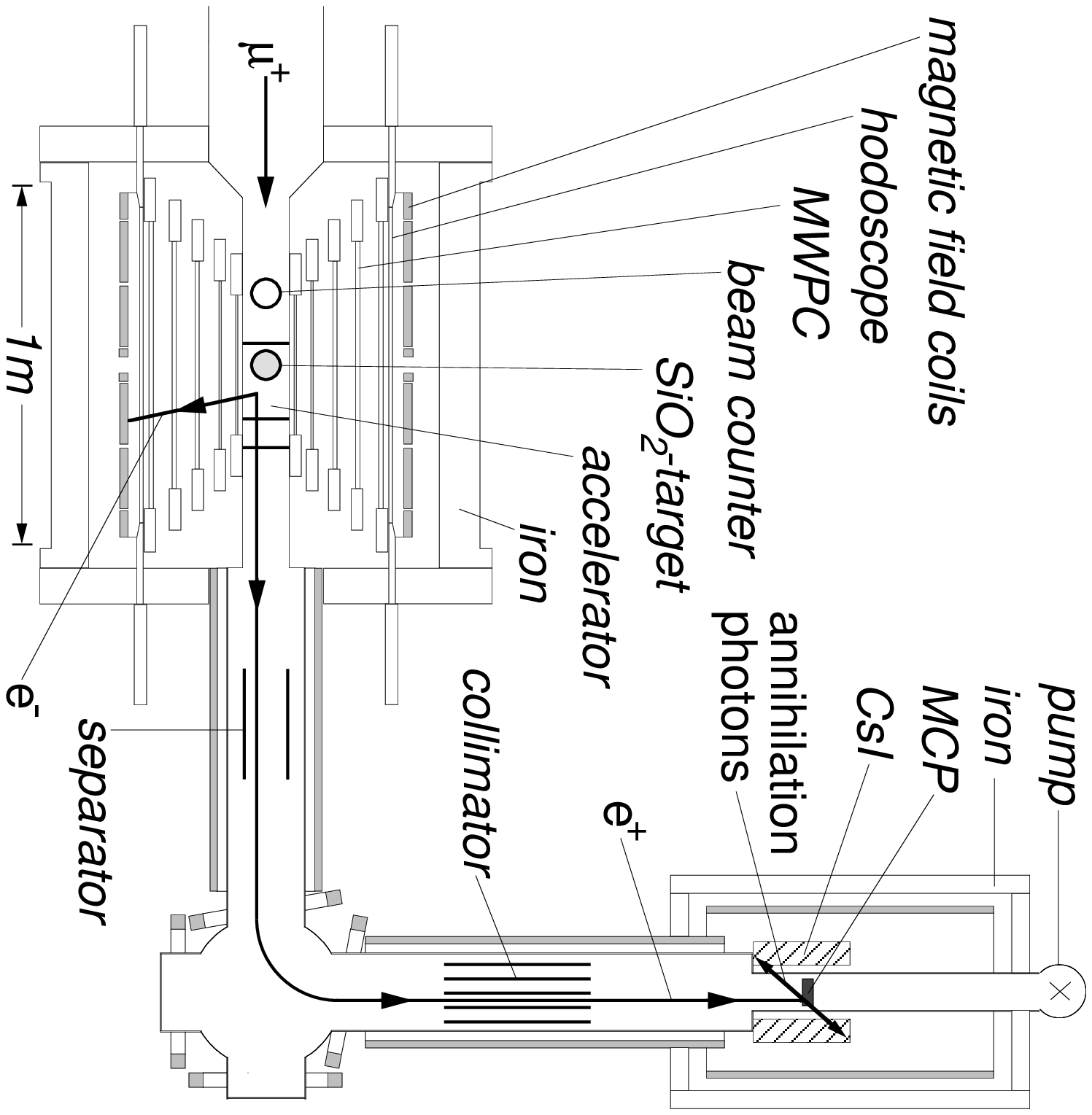,height=6.5cm,angle=90}
        }
\end{picture}
\end{minipage}
\end{figure}

\begin{figure}[ht]
\label{res_mmb}
\unitlength 1.0 cm
 \begin{minipage}{4.7cm}
 \begin{picture}(5.0,4.7)
  \centering{
   \hspace{-0.7 cm}
   \raisebox{0.0 cm}{
   \mbox{
   \psfig{figure=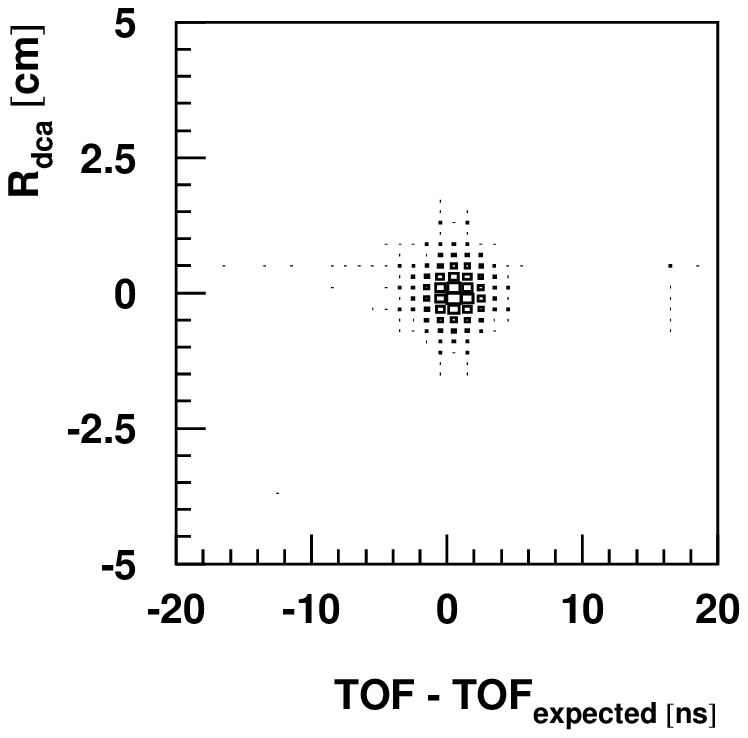,height=4.7cm}
        }
        }
             }
 \end{picture}
 \end{minipage}
 \hspace{0.3cm}
 \begin{minipage}{4.7cm}
 \begin{picture}(5.0,4.7)
  \centering{
   \raisebox{0.0 cm}{
   \mbox{
   \psfig{figure=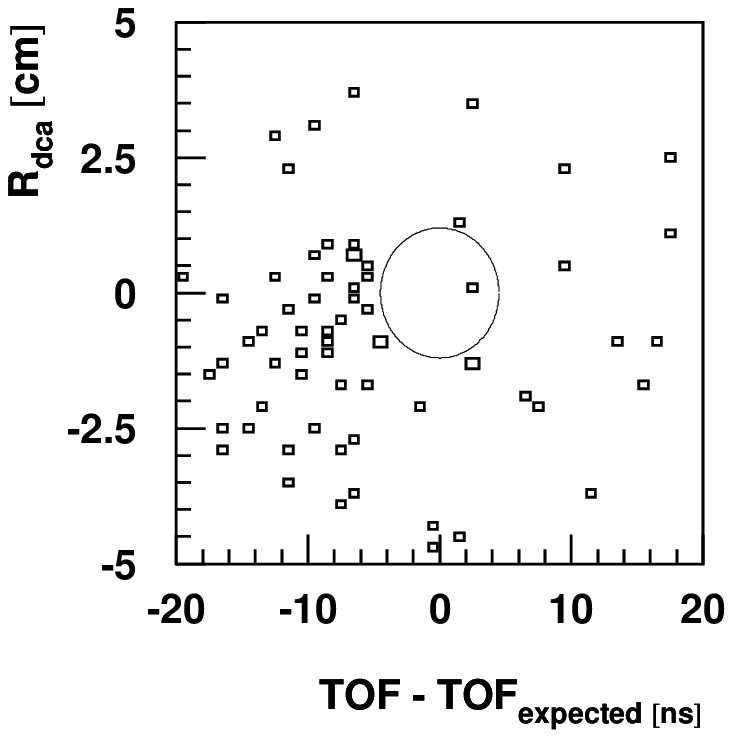,height=4.7cm}
        }
        }
             }
 \end{picture}
 \end{minipage}
 \centering\caption[]
        {Time of flight (TOF) and vertex quality for a muonium measurement 
        (left) and all data of 1996 while searching for antimuonium in 1996 
        (right). One event falls into the indicated 3 standard deviations area.
        }
\end{figure}

The relevant measurements were performed during in total 6 month 
distributed over 4 years.
The muonium production was monitored regularly 
by reversing all electric and magnetic fields of the instrument 
every five hours for a duration of 20 minutes.
Targets had to be replaced twice a week because of an observed 
deterioration of muonium production on a time scale of about one week. 
In the course of the experiment $5.7 \cdot 10^{10}$ muonium atoms 
were observed in the interaction volume for antimuonium decays.
There was one event which passed all required  criteria, i.e.
fell within a 99\% confidence interval of all relevant distributions
(Fig. \ref{res_mmb}).
The expected background due to accidental coincidences is 1.7(2) events.
The preliminary combination of all analyzed data 
results in 
an upper limit
for the conversion probability in a 0.1~T magnetic field of
${\rm P}_{{\rm M} \overline{{\rm M}}}\leq 8 \cdot 10^{-11}$.
For an assumed effective (V-A)x(V-A) type four fermion interaction
this corresponds to an upper limit
for the coupling constant of 
${\rm G}_{{\rm M} \overline{{\rm M}}}\leq 3\cdot 10^{-3} {\rm G}_{\rm F}$, 
where ${\rm G}_{\rm F}$, (90~\%~C.L.)
is the weak interaction Fermi coupling constant \cite{Meyer_97}.
This includes a correction for the magnetic field  \cite{Hou_95}.

This new result allows to rule out definitively 
a certain  ${\rm Z}_8$ model with more than 4 generations of particles 
\cite{Hou_96} 
and to
set a new lower limit of 2.6 TeV/c$^2$
on the masses of dileptonic gauge bosons in GUT models which is well 
beyond the value extracted from high energy Bhabha scattering \cite{Sasaki_94},
It can be further shown in the framework of minimal left right symmetric and
supersymmetric models that lepton number violating muon decay 
($\mu^+ \rightarrow e^+ + \nu_{\mu} + \overline{\nu}_e$) is not an option
for explaining the excess neutrino counts in the LSND neutrino
experiment at Los Alamos  \cite{Herczeg_97}.

\section{Magnetic anomaly of the muon}
The muon magnetic anomaly, i.e. the deviation of the muon's g-factor from the 
Dirac value 2, has been measured in a series of three 
experiments at CERN \cite{Farley_92} to an accuracy of 7~ppm. 
Although this quantity is determined, as in the case of the electron, 
mostly by photon and electron-positron fields,
the effects of heavier particles is enhanced compared
to the electron cases by the square of the mass ratio 
$(m_{\mu}/m_e)^2 \approx 4\cdot10^{4}$.
The contribution of the strong interaction is 60~ppm and the weak contribution
is expected to be 1.3~ppm. 

This summer a new experiment \cite{Roberts_97} 
was successfully brought into operation at the 
Brookhaven National Laboratory (BNL) in Upton, USA,
the ultimate goal of which is a 
determination of the muon 
\begin{table}[b]
\protect  \caption{
         Sensitivity to new physics of the new g-2 experiment at BNL.
         }
\vspace{0mm}
\protect \label{g2_new_physics}
{\small
  \begin{tabular}{lllll}
\hline
New Physics     &\multicolumn{3}{c}{Sensitivity}   &Other Experiments\\     
\hline
\hline\\
Muon substructure    & $\Lambda$& $\geq$ & 5~TeV      
& LHC similar\\
excited muon   & $m_{\mu^*}$& $\geq$ & 400~GeV  
& LEP II similar\\
W$^\pm$-boson substructure & $\Lambda$& $\geq$ & 400~GeV    
& LEP II $\sim$100-200~GeV \\
W$^\pm$ anomalous &$a_W$& $\geq$   & 0.02          
& LEP II $\sim$0.05, \\
\hspace*{3mm}magnetic moment&&&& LHC $\sim$0.2\\
Supersymmetry                 & $m_{\widetilde {W}}$& $\leq$   & 130~GeV       
& Fermilab p${\overline{\rm p}}$ similar\\
right handed W$^\pm_R$-bosons &$m_{W'}$& $\leq$    & 250~GeV       
& Fermilab p${\overline{\rm p}}$ similar\\
%
%
%
Muon el. dipole moment & $D_{\mu}$& $\leq$ & $4\cdot10^{-20}ecm$ 
                                        &       \\
\hline
  \end{tabular}
{\footnotesize
$^a$ for substructure $\Delta a_{\mu} \sim m^2_{\mu}/\Lambda^2$
}
}
 \end{table}
magnetic anomaly  to 0.35~ppm.
At this level of sensitivity numerous theories beyond the standard model
can be tested, respectively their parameters can be significantly restricted
(see Table 1) \cite{Roberts_97,Kinoshita_90}. 
The potential of this high precision experiment
reaches in part the one of high energy physics experiments and exceeds it even
some cases.

\section{Conclusions and Acknowledgments}

Precision experiments on muons and muonium can contribute to probe sensitively 
physics beyond the standard model in a complementary way to high energy 
physics approaches. Maybe ultimately they can contribute to resolve the
question of the difference in nature between muon and electron. 

This work was supported by the German Bundesminister f\"ur Bildung und 
Forschung and by a NATO research grant.

\end{document}